\documentclass[10pt]{article}

\usepackage{amstext}
\usepackage{amsmath}
\usepackage{amssymb}
\usepackage{a4}

\newtheorem{theorem}{Theorem}[section]
\newtheorem{lemma}[theorem]{Lemma}
\newtheorem{proposition}[theorem]{Proposition}

\newcommand{\Proof}{\par\noindent{\em Proof. }}
\newcommand{\eop}{\nopagebreak\hspace*{\fill}$\Box$\smallskip}

\def\XXint#1#2#3{{\setbox0=\hbox{$#1{#2#3}{\int}$}
     \vcenter{\hbox{$#2#3$}}\kern-.5\wd0}}

\newcommand{\N}{\Bbb N}
\newcommand{\Z}{\Bbb Z}
\newcommand{\R}{\Bbb R}
\newcommand{\C}{\Bbb C}

\newcommand{\eps}{\varepsilon}

\newcommand{\supp}{\operatorname{supp}}
\newcommand{\trace}{\operatorname{trace}}
\newcommand{\dist}{\operatorname{dist}}

\begin{document}

\begin{center}
\begin{Large}
{\bf Localized spectral asymptotics for boundary value problems and correlation effects in the free Fermi gas in general domains}
\end{Large}
\end{center}

\begin{center}
\begin{large}
Bernd Schmidt\\
\end{large}
\begin{small}
Zentrum Mathematik, Technische Universit{\"a}t M{\"u}nchen\\ 
Boltzmannstr.\ 3, 85747 Garching, Germany\\ 
{\tt schmidt@ma.tum.de}
\end{small}
\end{center}

\begin{center}
\today
\end{center}
\bigskip

\begin{abstract}
We rigorously derive explicit formulae for the pair correlation function of the ground state of the free Fermi gas in the thermodynamic limit for general geometries of the macroscopic regions occupied by the particles and arbitrary dimension. As a consequence we also establish the asymptotic validity of the local density approximation for the corresponding exchange energy. At constant density these formulae are universal and do not depend on the geometry of the underlying macroscopic domain. In order to identify the correlation effects in the thermodynamic limit, we prove a local Weyl law for the spectral asymptotics of the Laplacian for certain quantum observables which are themselves dependent on a small parameter under very general boundary conditions. 
\end{abstract}

\begin{small}

\end{small}

\section{Introduction and main results}

A classical result of Wigner and Seitz provides an explicit formula for the ``exchange hole'' in ground states of the free electron gas in a cubical box for a large number of particles in the thermodynamic limit, \cite{WignerSeitz}: Although the quantum mechanical Hamiltonian does not contain coupling terms between different particles, due to the Pauli principle, i.e., the requirement that the quantum mechanical wave function be antisymmetric with respect to particle exchange, the statistical distribution of the individual electrons in $\Omega$ is not independent. This phenomenon can be measured in terms of the autocorrelation function $P(x, y)$ which is given by the difference of the two-body density $\rho_2(x + \frac{y}{2},x - \frac{y}{2})$ and a statistically independent superposition of the one-body densities $\rho_1(x + \frac{y}{2})$ and $\rho_1(x - \frac{y}{2})$, see Section \ref{sec:Fermi-gas} for details.  

If the number $N$ of electrons in $\Omega = (0, L)^3$ tends to infinity at constant microscopic density $\bar{\rho} = N/L^3$, the Wigner-Seitz formula gives an asymptotic expression for the autocorrelation function:  
\begin{align}\label{eq:Wigner-Seitz} 
  P(x, y) 
  \approx - \frac{\bar{\rho}}{4} \left( \frac{3 ( \sin (p_F |y|) 
  - p_F |y| \cos (p_F |y|) ) }{(p_F |y|)^3} \right)^2, 
\end{align} 
where $p_F = (3 \pi^2 \bar{\rho})^{1/3}$ is the Fermi-momentum of the free electron gas with density $\bar{\rho}$ (in atomic units). 

Corresponding to the exchange hole in the particle density there is the ``exchange energy'' $E_{\rm x}(A)$ which measures the difference in the Coulomb energy of the true ground state in the region $A \subset \Omega$ as compared to a statistically independent particle distribution, cf.\ Section \ref{sec:Fermi-gas} for details. Results of Dirac, Bloch, Slater and G{\'a}sp{\'a}r (cf.\ \cite{Dirac,Bloch,Slater,Gaspar}) relate this exchange energy to the ``local density approximation'' 
\begin{align}\label{eq:Dirac-Bloch} 
  E_{\rm x}(A) 
  \approx - c_{\rm x} \int_{A} \rho_1^{4/3}(x) \, dx  
\end{align} 
with $c_{\rm x} = \frac{3}{4}(\frac{3}{\pi})^{1/3}$. 

While these results are classical by now and can be found in standard text books on density functional theory as, e.g., by Eschrig \cite{Eschrig} or Szabo and Ostlund \cite{SzaboOstlund}, rigorous convergence proofs seem to have been obtained only much later. The book of Thirring \cite{Thirring} and the articles of Bach \cite{Bach} and Graf and Solovej \cite{GrafSolovej} give rigorous estimates on the exchange energy (not only for the free system). However, the first complete convergence proof for the pair correlation function under both Dirichlet and periodic boundary conditions has been obtained by Friesecke in \cite{Friesecke}. In this paper the author also gives a thorough analysis of boundary layer effects and sharp error estimates. However, it appears that all formal derivations and rigorous results depend heavily on the explicit knowledge of the eigenfunctions of the one-body Laplacian on $\Omega$. In particular, these calculations do not seem to be easily extendable to general domains $\Omega$. 

This situation is reminiscent of -- and as we will see in fact closely related to -- the problem of finding a general asymptotic law for the distribution of the eigenvalues of the Laplacian on a general domain $\Omega \subset \R^n$. For $\Omega = (0,L)^3$ it had been well known that the number of eigenvalues $\lambda_1 \le \lambda_2 \le \ldots$ of the Laplacian with Dirichlet or Neumann boundary values asymptotically satisfies 
$$ \lambda_k 
   \sim 6^{2/3} \pi^{4/3}L^{-2} k^{2/3} 
$$
for large $k$, in the sense that the ratio between the left and the right hand side converges to $1$ for $k \to \infty$. On physical grounds, at the beginning of the last century Lorentz and Sommerfeld had conjectured that the same formula holds true for general domains $\Omega \subset \R^3$. So, in particular, the asymptotic distribution of the eigenvalues should depend on $\Omega$ only through the volume $|\Omega|$. This conjecture was proved not much later by Weyl, cf.\ \cite{Weyl:11,Weyl:12}. In general dimension $n$, his result reads 
\begin{align}\label{eq:Weyl} 
  \lim_{k \to \infty} \frac{\lambda_k^{n/2}}{k} = \frac{(4 \pi)^{n/2} \Gamma (\frac{n}{2} + 1)}{|\Omega|}, 
\end{align} 
a formula now known as ``Weyl's law''. It holds true, e.g., for the Dirichlet Laplacian on general domains $\Omega \subset \R^n$ as well as for the Neumann Laplacian under suitable regularity assumptions, which are in particular satisfied if $\partial \Omega$ is Lipschitz. See, e.g., the recent article of Netrusov and Safarov \cite{NetrusovSafarov} on Weyl's law for very general domains. 

Denoting the number of eigenvalues less than or equal to some $\lambda > 0$ by $N(\lambda)$ and noting that the volume of the ball $B_r$ of radius $r$ about $0$ is $|B_r| = \pi^{n/2} r^n/\Gamma (\frac{n}{2} + 1)$, we see that Weyl's formula can equivalently be written as 
\begin{align}\label{eq:Weyl-alternativ} 
  (2 \pi)^{n} N(\lambda) 
  \sim | \Omega \times B_{\sqrt{\lambda}} |, 
\end{align} 
for $\lambda \to \infty$ and thus relates the number of eigenvalues $\le \lambda$ to the classical phase space volume of $\Omega \times B_{\sqrt{\lambda}}$. 

Now if $(u_k)$ is an orthonormal series of eigenfunctions corresponding to $(\lambda_k)$, then obviously $N(\lambda) = \sum_{k \le \lambda} (u_k, u_k)_{L^2}$. Correspondingly, a local Weyl law gives, generally speaking, the asymptotics of localized quantities of the form $\sum_{k \le \lambda} (u_k, A u_k)_{L^2}$, where $A$ is a suitable operator. Results in this direction have been obtained for pseudodifferential operators $A$ of degree $0$ for instance when the $u_k$ are eigenfunctions of Schr{\"o}dinger operators $-\Delta + V$ with smooth potentials $V$ under growth assumptions $V(x) \to \infty$ as $|x|\to \infty$, see, e.g., Zelditch's ``Szeg{\"o} limit theorems'' \cite{Zelditch} and also compare the lecture notes \cite{EvansZworski} of Evans and Zworski. The volume term $|\Omega \times B_{\sqrt{\lambda}}|$ on the right hand side of \eqref{eq:Weyl-alternativ} then has to be replaced with the localized term 
$$ \int_{\{\xi^2 + V(x) \le \sqrt{\lambda} \}} a(x, \xi) \, dx \, d\xi, $$
where $a$ denotes the symbol of $A$. 

Starting with the seminal paper of Shnirelman \cite{Shnirelman} there have been asymptotic results even for the individual terms $(u_k, A u_k)_{L^2}$ for the eigenfunctions $u_k$ of the Laplacian on Riemannian manifolds under the assumption that the geodesic flow on the unit cotangent bundle of $M$ be ergodic. For Euklidean domains with smooth boundaries and under ergodicity assumptions of he correponding billiard system such results for the (free) Laplacian with Dirichlet boundary values are obtained in the fundamental contribution by G{\'e}rard and Leichtnam \cite{GerardLeichtnam}. In contrast, our main focus will lie on the asymptotics of the Ces{\`a}ro means $N^{-1} \sum_{k = 1}^N (u_k, A u_k)_{L^2}$ for general domains $\Omega$. Indeed, radially averaged limits of these quantities with polyhomogeneous $A$ have been considered previously in \cite{GerardLeichtnam}, too. Note, however, that the operators $A$ to be investigated below do not introduce radial averaging of the $u_k$. Moreover, they will themselves contain a small length scale depending on $N$. Also, our results will hold true even for very rough boundaries $\partial \Omega$. 

More precisely, our first main result will be a local Weyl law which involves two small scales. Let $\Omega \subset \R^n$ be a bounded domain. We consider  boundary value problems for the Laplacian on $\Omega$ subject to the boundary condition ${\cal B}u = 0$ such that $-\Delta$ has a pure point spectrum consisting of the eigenvalues $\lambda_1 \le \lambda_2 \le \cdots$ with corresponding eigenfunctions $u_k$ that form an orthonormal basis of $L^2(\Omega)$: 
\begin{align}\label{eq:uk-glg}
  - \Delta u_k = \lambda_k u_k \mbox{ in } \Omega, 
  \qquad 
  {\cal B} u_k = 0 \mbox{ on } \partial \Omega, 
\end{align}
such that the sequence $(\lambda_k)$ of eigenvalues obeys Weyl's law \eqref{eq:Weyl}. Note that, e.g., this condition is satisfied for the Dirichlet Laplacian on domains $\Omega$ as well as for the Neumann Laplacian on $\Omega$ if $\partial \Omega$ is sufficiently regular, say Lipschitz (see, e.g., \cite{NetrusovSafarov}). Define 
\begin{align}\label{eq:gamma-defi}
  \gamma = \gamma(\Omega) = \frac{2 \pi^{1/2} \Gamma^{1/n} (\frac{n}{2} + 1)}{|\Omega|^{1/n}} 
\end{align}
and the sequence $h_k = \lambda_k^{-1/2}$, so that Weyl's formula \eqref{eq:Weyl} reads $\lim_{k \to \infty} k^{1/n} h_k = \gamma^{-1}$. 

The quantities $h_k := \lambda_k^{-1/2} \to 0$ as $k \to \infty$ describe the natural scale for measuring the oscillatory wave-length in $u_k$. On the other hand, after rescaling to a fixed macroscopic domain $\Omega$, the autocorrelation function for a typical ground state of the free Fermi gas is given (say, for simplicity, in the spinless case) by  
\begin{align}\label{eq:Auto-gleich}
  P_{N} (x, y) 
   = -\frac{1}{2} \left| N^{-1} \sum_{k = 1}^N u_k \left( x + \frac{N^{-1/n} y}{2} \right) \, 
      \overline{u_k}\left( x - \frac{N^{-1/n} y}{2} \right) \right|^2. 
\end{align} 
This introduces a second small length-scale $N^{-1/n}$. The determination of the sum in \eqref{eq:Auto-gleich} is non-trivial due to the subtle interplay of the two small parameters: By Weyl's law, $|\nabla u_k(x)| |\frac{N^{-1/n} y}{2}|$ scales like $h_k^{-1} N^{-1/n}$ and thus like $(k/N)^{1/n}$, so the terms $u_k(x \pm \frac{N^{-1/n} y}{2})$ are expected to be close to $u_k(x)$ for $k \ll N$ and highly oscillating for $k \gg N$, but to yield non-negligible contributions to $P_N(x, y)$ for $k$ of the same order as $N$, where the summation is truncated. 

We denote by $(v, w) = (v, w)_{L^2} = \int_{\R^n} \overline{v} \, w$ the $L^2(\R^n)$-inner product on of functions two $v$ and $w$. Deferring the precise definition of the pseudodifferential operator $a^{\rm w}(x, hD)$ obtained by the semiclassical Weyl quantization of a symbol $a = a(x, \xi)$ of class $S$ to Section \ref{sec:Weyl}, the result reads as follows.   
\begin{theorem}\label{theo:conv-to-trace} 
Suppose $\Omega \subset \R^n$ is a bounded domain and $(u_k)$ is an orthonormal sequence of eigenfunctions for the boundary value problem \eqref{eq:uk-glg}, extended by $0$ outside $\Omega$, with corresponding eigenvalues $\lambda_1 \le \lambda_2 \le \ldots$ that satisfy Weyl's law. If $a \in S$, then 
$$ \lim_{N \to \infty} N^{-1} \sum_{k = 1}^N ( u_k, a^{\rm w}(x, N^{-1/n}D) u_k ) 
   = (2\pi)^{-n} \int_{\Omega \times B_{\gamma}} a(x, \xi) \, dx \, d\xi $$ 
for $\gamma$ as defined in \eqref{eq:gamma-defi}. 
\end{theorem}

Note that $\gamma^n = \frac{(2\pi)^n}{|\Omega \times B_1|}$ and hence 
\begin{align}\label{eq:W-Mass} 
  (2\pi)^{-n} |\Omega \times B_{\gamma}| 
  = 1. 
\end{align}
So as a functional of $a$, the limiting expression is nothing but the uniform distribution on $\Omega \times B_{\gamma}$. 

The starting point for our investigation of correlation effects will be a convergence result for the so-called one-body matrix, which, rescaled to fixed $\Omega$, reads 
\begin{align}\label{eq:QN-defi}
  Q_N(x, y) 
  = N^{-1} \sum_{k = 1}^N u_k \left( x + \frac{N^{-1/n} y}{2} \right) \, 
  \overline{u_k}\left( x - \frac{N^{-1/n} y}{2} \right). 
\end{align} 
Observing that the Wigner transformation relates this expression to averaged  quantum observables, by applying Theorem \ref{theo:conv-to-trace} we will be able to obtain the following convergence result. 
\begin{theorem}\label{theo:one-body-matrix} 
Under the conditions of Theorem \ref{theo:conv-to-trace}, the one-body matrices $Q_{N}$ in \eqref{eq:QN-defi} converge to 
\begin{align*}
  Q(x, y) 
  = \frac{2^{n/2} \Gamma(\frac{n}{2} + 1) J_{n/2}(\gamma |y|)}{|\Omega|(\gamma |y|)^{n/2}} \chi_{\Omega}(x)
\end{align*} 
strongly in $L^2(\R^n \times \R^n)$ and boundedly in measure on compact subsets of $\Omega \times \R^n$, where $J_{n/2}$ denotes the Bessel function of the first kind of order $\frac{n}{2}$ and $\gamma$ is defined by \eqref{eq:gamma-defi}. 
\end{theorem}
Since $\gamma^n = \frac{(2\pi)^n}{|\Omega \times B_{1}|}$, $\chi_{B_{\gamma}} = \chi_{B_1}(\gamma^{-1}\cdot)$ and the Fourier transform of the characteristic function of the unit ball is given by $\widehat{\chi_{B_1}}(y) = (2 \pi)^{n/2} |y|^{-n/2} J_{n/2}(|y|)$, the term on the right hand side is nothing but 
$$ Q(x, y) 
   = \frac{\widehat{\chi_{B_{\gamma}}}(y)}{|\Omega \times B_{\gamma}|} \chi_{\Omega}(x).$$ 
Also note that, although our assumptions on the boundary $\partial \Omega$ are very weak, we do obtain strong $L^2$-convergence on the whole phase space $\R^n \times \R^n$. As expected, away from the boundary the convergence is even stronger. 

As will be detailed in Section \ref{sec:Fermi-gas}, for a ground state of the free Fermi gas with $m$ spin states, which is given as a Slater determinant of the (rescaled) single particle wave functions $\psi_1, \ldots, \psi_N \in L^2(\Omega \times {\cal S})$, the (rescaled) one-body matrix is given by 
\begin{align}\label{eq:QN-spin-defi}
  Q^{\cal S}_{N}(x, y, s_1, s_2) 
  = N^{-1} \sum_{i = 1}^N \psi_{i} \left( x + \frac{N^{-1/n} y}{2}, s_1 \right) 
  \overline{\psi_{i}} \left( x - \frac{N^{-1/n} y}{2}, s_2 \right). 
\end{align} 
Throughout we will assume that the $\psi_i$ (extended by $0$ outside $\Omega$) satisfy the boundary condition ${\cal B} \psi_i(\cdot, s) = 0$ on $\partial \Omega$ for the bounded domain $\Omega$ so that the set of eigenvalues $(u_k)$ corresponding to \eqref{eq:uk-glg} satisfies the assumptions of Theorem \ref{theo:conv-to-trace}. The corresponding convergence result reads 
\begin{theorem}\label{theo:one-body-matrix-spin} 
Let $s_1, s_2 \in {\cal S}$. The rescaled one-body matrices $Q_{N}(\cdot,\cdot,s_1,s_2)$ in \eqref{eq:QN-spin-defi} corresponding to determinantal ground states of the $N$-body free Fermi gas converge to 
\begin{align*}
  Q^{\cal S}(x, y, s_1, s_2) 
  = \frac{2^{n/2} \Gamma(\frac{n}{2} + 1) J_{n/2}(p_F |y|) \delta_{s_1 s_2}}{m |\Omega|(p_F |y|)^{n/2}} \chi_{\Omega}(x)
\end{align*} 
strongly in $L^2(\R^n \times \R^n)$ and boundedly in measure on compact subsets of $\Omega \times \R^n$, where $p_F = \gamma m^{-1/n}$. 
\end{theorem}
We note that $p_F$ is the ``Fermi momentum'' of the original unrescaled problem: By Weyl's law, 
\begin{align}\label{eq:Fermi-momentum}
  p_F 
  = \gamma m^{-1/n} 
  = \lim_{N \to \infty} N^{-1/n} \gamma \lfloor m^{-1} N \rfloor^{1/n} 
  = \lim_{N \to \infty} N^{-1/n} \lambda_{\lfloor N/m \rfloor}^{1/2}. 
\end{align}

As a direct consequence of these theorems, we obtain a generalization of the Wigner-Seitz formula \eqref{eq:Wigner-Seitz} to general shapes of domains in arbitrary dimensions, which for the sake of clarity we first state for the spinless mere boundary value problem. 
\begin{theorem}\label{theo:wigner-seitz} 
Under the conditions of Theorem \ref{theo:conv-to-trace}, the autocorrelation functions $P_{N}$ in \eqref{eq:Auto-gleich} converge to 
$$ P(x, y) 
   = - \frac{2^{n-1} \Gamma^2(\frac{n}{2} + 1) J^2_{n/2}(\gamma |y|)}{|\Omega|^2(\gamma |y|)^{n}} \chi_{\Omega}(x) $$ 
strongly in $L^1(\R^n \times \R^n)$ and boundedly in measure on compact subsets of $\Omega \times \R^n$.
\end{theorem}
Moreover, we obtain strong convergence of the so-called one-body density 
\begin{align}\label{eq:one-body-density}
  \rho_{1,N} (x) 
  := N^{-1} \sum_{k = 1}^N |u_k(x)|^2. 
\end{align} 
to the uniform distribution on $\Omega$. (Vague convergence of the one-body density for the Dirichlet eigenfunctions on smooth domains is in fact inherent already in \cite{GerardLeichtnam}.) 
\begin{theorem}\label{theo:one-body-conv} 
Under the conditions of Theorem \ref{theo:conv-to-trace}, the one-body densities $\rho_{1,N}$ in \eqref{eq:one-body-density} converge to $\bar{\rho} = \frac{\chi_{\Omega}}{|\Omega|}$ strongly in $L^1$ and boundedly in measure on compact subsets of $\Omega$.  
\end{theorem}

For general spin systems we obtain the following generalized formula for the exchange hole: Let 
\begin{align}\label{eq:AutoSpin-gleich} 
  P^{\cal S}_{N}(x, y) 
  = - \frac{1}{2} \sum_{s \in {\cal S}^2}  
  \left| N^{-1} \sum_{i = 1}^N \psi_{i} \left( x + \frac{N^{-1/n} y}{2}, s_1 \right) 
  \overline{\psi_{i}} \left( x - \frac{N^{-1/n} y}{2}, s_2 \right) \right|^2 
\end{align} 
(cf.\ Section \ref{sec:Fermi-gas} for a derivation). 

\begin{theorem}\label{theo:wigner-seitz-spin}
The rescaled autocorrelation functions $P^{\cal S}_{N}$ in \eqref{eq:AutoSpin-gleich} corresponding to determinantal ground states of the $N$-body free Fermi gas converge to 
$$ P^{\cal S}(x, y) 
   = - \frac{2^{n-1} \Gamma^2(\frac{n}{2} + 1) J_{n/2}^2(p_F |y|)}{m |\Omega|^2 (p_F|y|)^n} 
   \chi_{\Omega}(x) $$ 
strongly in $L^1(\R^n \times \R^n)$ and boundedly in measure on compact subsets of $\Omega \times \R^n$, where $p_F$ is the Fermi-momentum from \eqref{eq:Fermi-momentum}.
\end{theorem}

As $J_{3/2}(r) = \sqrt{\frac{2}{\pi}} \frac{\sin r -r \cos r}{r^{3/2}}$, in $n = 3$ dimensions the limiting expression reduces to 
\begin{align*}  
  P^{\cal S}(x, y) 
  = - \frac{\bar{\rho}^2}{2m} \left( \frac{ 3(\sin (p_F |y|) - p_F |y| \cos (p_F |y|)) }
    {(p_F |y|)^3} \right)^2 
\end{align*} 
on $\Omega$ with $p_F = \left( \frac{6 \pi^{2} \bar{\rho}}{m} \right)^{1/3}$, $\bar{\rho} = |\Omega|^{-1}$. For the electron gas with $m = 2$ we indeed recover the Wigner-Seitz formula \eqref{eq:Wigner-Seitz}. 

The statement for the one-body density in the general Fermi gas is analogous to the spinless case. As justified in Section \ref{sec:Fermi-gas} we let  
\begin{align}\label{eq:one-body-density-spin}
  \rho^{\cal S}_{1,N} (x) 
  := N^{-1} \sum_{s \in {\cal S}} \sum_{i = 1}^N |\psi_{i}(x, s)|^2. 
\end{align} 
\begin{theorem}\label{theo:one-body-conv-spin} 
The one-body-densities $\rho^{\cal S}_{1,N}$ in \eqref{eq:one-body-density-spin} corresponding to determinantal ground states of the $N$-body free Fermi gas converge to $\bar{\rho} = \frac{\chi_{\Omega}}{|\Omega|}$ strongly in $L^1$ and boundedly in measure on compact subsets of $\Omega$.  
\end{theorem}

Thanks to the strong convergence of $P^{\cal S}_N$ proved in Theorem \ref{theo:wigner-seitz-spin}, we can also generalize the Dirac-Bloch formula \eqref{eq:Dirac-Bloch} for the exchange energy in the local density approximation to arbitrary domains. It turns out that, for fixed $m$, the constant $c_{\rm x}$ computed for the box $(0,L)^3$ is in fact universal in three dimensions. 

Recall the definitions of $P^{\cal S}_N$ and $\rho^{\cal S}_{1,N}$ from \eqref{eq:AutoSpin-gleich} resp.\ \eqref{eq:one-body-density-spin} and set  
\begin{align}\label{eq:Ex-defi}
  E^{\cal S}_{{\rm x},N}(A) = \int_{A \times \R^n} \frac{P^{\cal S}_N(x, y)}{|y|}. 
\end{align} 
\begin{theorem}\label{theo:Ex-LDA} 
Let $n = 3$. For any $A \subset\subset \Omega$ 
$$ \lim_{N \to \infty} E^{\cal S}_{{\rm x}, N} (A) 
   = \lim_{N \to \infty} c_{\rm x}  \int_{A} (\rho^{\cal S}_{1,N})^{4/3} 
   = c_{\rm x}  |\Omega|^{-4/3} |A|, $$ 
where $c_{\rm x} = 3(\frac{3}{32 \pi m})^{1/3}$. 
\end{theorem}
For the electron gas with $m = 2$ and $c_{\rm x} = \frac{3}{4}(\frac{3}{\pi})^{1/3}$ we obtain that formula \eqref{eq:Dirac-Bloch} is indeed valid for general domains $\Omega$.  

Having stated all our main results, we end this section with a brief outline of the following chapters. Section \ref{sec:Fermi-Weyl} contains the basic material on the physics of the free Fermi gas and on Weyl quantization, which will be needed throughout this paper. In Section \ref{sec:loc-spec} we prove the localized Weyl formula of Theorem \ref{theo:conv-to-trace} for boundary value problems. This will be applied in Section \ref {sec:averaged-ev} to prove Theorem \ref{theo:one-body-matrix}, which encodes the pair correlation effects in the Ces{\`a}ro averaged system of eigenfunctions. Theorems \ref{theo:wigner-seitz} and \ref{theo:one-body-conv} are straightforward corollaries. Systems with spins will then be analyzed in Section \ref{sec:Fermi-gas-corr}. First Theorem \ref{theo:one-body-matrix-spin} will be proved by reduction to the spinless case. Then similarly as in the spinless case, Theorems \ref{theo:wigner-seitz-spin}, \ref{theo:one-body-conv-spin} and \ref{theo:Ex-LDA} will be direct consequences of this result.

\section{The free Fermi gas and Weyl's calculus}\label{sec:Fermi-Weyl}

This chapter serves to collect some background material: firstly, on the quantum mechanical description of a Fermi gas and, secondly, on Weyl's pseudodifferential calculus for quantum observables.  

\subsection{Quantum mechanics of the free Fermi gas}\label{sec:Fermi-gas}

In this paragraph we will briefly recall some basic material on the correlation function of the free Fermi gas, in particular, its connection to the one- and two-body densities of the associated quantum mechanical wave function. General references are, e.g., \cite{Eschrig,SzaboOstlund}. 

A Fermi gas of $N$ particles in a region $\Omega_N \subset \R^n$ is described by an $N$-body quantum mechanical wave function $\psi : (\Omega_N \times {\cal S})^N \to \C$, $(x_1, s_1, x_2, s_2, \ldots, x_N, s_N) \mapsto \psi(x_1, \ldots, s_N)$ which is $L^2$-normalized, i.e., 
$$ \sum_{s \in {\cal S}^N} \int_{\Omega_N^{N}} 
   |\psi(x_1, s_1, x_2, s_2, \ldots, x_N, s_N)|^2 
   \, dx_1 \cdots dx_N $$ 
and satisfies the Pauli principle, i.e., is antisymmetric with respect to particle exchange. Here ${\cal S} = \{\sigma_1, \ldots,\sigma_m\}$ denotes the set of spin variables, e.g., $\{\sigma_1,\sigma_2\} =$ $\{\uparrow,\downarrow\}$ for an electron gas with two states of spin: ``up'' and ``down''. In addition, $\psi$ is subject to suitable boundary conditions, e.g., Dirichlet conditions for particles bound to $\Omega_N$ by an infinitely deep potential well.

Particular wave functions are given by the so-called Slater determinants 
\begin{align}\label{eq:Slater}
 \psi(x_1, \ldots, s_N) 
   = \frac{1}{\sqrt{N!}} \det 
     \begin{pmatrix} 
        \psi^{(\Omega_N)}_1(x_1, s_1) & \cdots & \psi^{(\Omega_N)}_1(x_N, s_N) \\ 
        \vdots & & \vdots \\ 
        \psi^{(\Omega_N)}_N(x_1, s_1) & \cdots & \psi^{(\Omega_N)}_N(x_N, s_N) 
     \end{pmatrix}. 
\end{align} 
Here the $\psi^{(\Omega_N)}_i$ are $N$ orthonormal single-particle wave functions. If $\lambda_1 \le \lambda_2 \le \ldots$ denotes the eigenvalues of the Laplacian on $\Omega_N$ and $(u_k)$ a corresponding set of orthonormal eigenfunctions subject to the boundary condition ${\cal B} u_k = 0$ on $\partial \Omega$, then a particular orthonormal basis of the single-particle wave functions is given by 
$$ u_1 \otimes \delta_{\sigma_1}, \ldots, u_1 \otimes \delta_{\sigma_m}, 
   u_2 \otimes \delta_{\sigma_1}, \ldots, u_2 \otimes \delta_{\sigma_m}, 
   u_3 \otimes \delta_{\sigma_1}, \ldots $$ 
with corresponding energies $\mu_1 = \ldots = \mu_m = \lambda_1$, $\mu_{m+1} = \ldots = \mu_{2m} = \lambda_2$, \ldots.

The wave function $\psi$ is a ground state for the free Fermi gas if it minimizes the kinetic energy 
$$ \frac{1}{2} \sum_{i = 1}^N \sum_{s \in {\cal S}^N} \int_{\Omega_N^N} 
   |\nabla_{x_i} \psi(x_1, \ldots, s_N)|^2 
   \, dx_1 \cdots dx_N. $$
It is well known (and not hard to prove) that a Slater determinant $\psi$ is a ground state if and only if $\psi$ is the Slater determinant of $N$ orthonormal single-particle wave functions $\psi^{(\Omega_N)}_1, \ldots, \psi^{(\Omega_N)}_N$ corresponding to the $N$ lowest eigenvalues of the one-body problem. 

For given $\psi$, the one-body density is obtained by fixing a point $x \in \Omega_N$ integrating over the remaining spatial variables and all of the spin variables: 
\begin{align}\label{eq:one-body}
   \rho_1^{(\Omega_N)}(x) 
   := N \sum_{s \in {\cal S}^N} \int_{\Omega_N^{N-1}} 
   |\psi(x, s_1, x_2, s_2, \ldots, x_N, s_N)|^2 
   \, dx_2 \cdots dx_N, 
\end{align}
the two-body density correspondingly by fixing two spatial points $x$ and $\tilde{x}$ and integrating over the remaining spatial variables and all of the spin variables: 
\begin{align}\label{eq:two-body}
  \rho_2^{(\Omega_N)}(x, x') 
  := {N \choose 2} \sum_{s \in {\cal S}^N} \int_{\Omega_N^{N-2}} 
  |\psi(x, s_1, x', s_2, x_3, s_3, \ldots, x_N, s_N)|^2 
  \, dx_3 \cdots dx_N. 
\end{align}
The pair correlation function $P^{(\Omega_N)}$ then measures the difference of the two-body density as compared to a statistically independent superposition of the single particle densities: 
$$ P^{(\Omega_N)}(x, y) 
   := \rho_2^{(\Omega_N)} \left( x + \frac{y}{2},x - \frac{y}{2} \right) 
     - \frac{1}{2} \rho_1^{(\Omega_N)} \left( x + \frac{y}{2} \right) \rho_1^{(\Omega_N)} \left( x - \frac{y}{2} \right). $$

If $\psi$ is a Slater determinant as in \eqref{eq:Slater} it is well known (and easy to show) that 
\begin{align*}
  \rho_1^{(\Omega_N)}(x) 
  &= \sum_{s \in {\cal S}} \sum_{i = 1}^N |\psi_{i}^{(\Omega_N)}(x, s)|^2 
\end{align*}
and 
\begin{align*}
  \rho_2^{(\Omega_N)}(x, x') 
  &= \sum_{s \in {\cal S}^2} \frac{1}{2} \sum_{1 \le i,j \le N} 
     |\psi_{i}^{(\Omega_N)}(x, s_1)|^2 |\psi_{j}^{(\Omega_N)}(x', s_2)|^2 \\ 
  &\hspace{2.5cm} - \psi_{i}^{(\Omega_N)}(x, s_1) \overline{\psi_{j}^{(\Omega_N)}}(x, s_1) 
     \psi_{j}^{(\Omega_N)}(x', s_2) \overline{\psi_{i}^{(\Omega_N)}}(x', s_2) 
\end{align*}
and, consequently, 
\begin{align*}
  \rho_2^{(\Omega_N)}(x, x') - \frac{1}{2} \rho_1^{(\Omega_N)}(x) \rho_1^{(\Omega_N)}(x') 
  = - \frac{1}{2} \sum_{s \in {\cal S}^2}  
     \left| \sum_i \psi_{i}^{(\Omega_N)}(x, s_1) \overline{\psi_{i}^{(\Omega_N)}}(x', s_2) \right|^2. 
\end{align*}
It will be convenient to also introduce the one body matrix expression 
\begin{align*}
   Q^{(\Omega_N)}(x, y,s,s') 
   &:= N \sum_{s_2,\ldots,s_N \in {\cal S}} \int_{\Omega_N^{N-1}} 
   \psi \left( x + \frac{y}{2}, s, x_2, s_2, \ldots, x_N, s_N \right) \\ 
   &\hspace{3cm} \cdot \overline{\psi} \left( x - \frac{y}{2}, s', x_2, s_2, \ldots, x_N, s_N \right) 
   \, dx_2 \cdots dx_N, 
\end{align*}
so that 
\begin{align*}
   Q^{(\Omega_N)}(x, y,s,s') 
   = \sum_i \psi_{i}^{(\Omega_N)}(x, s) \, \overline{\psi_{i}^{(\Omega_N)}}(x', s').  
\end{align*}

Now in order to investigate the convergence properties of $Q^{(\Omega_N)}$ and hence $P^{(\Omega_N)}$ for $y$ of order one in the thermodynamic limit $N \to \infty$ and $\Omega_N = N^{1/n} \Omega \to \R^n$, we rescale to a fixed macroscopic region $\Omega$. Noting that the eigenfunctions scale like $\psi_{i}(x) = \psi_{i}^{(\Omega)}(x) = N^{1/2} \psi_{i}^{(\Omega_N)}(N^{1/n}x)$, we arrive at $Q^{\cal S}_N : \R^n \times \R^n \times {\cal S} \times {\cal S} \to \R$, $P^{\cal S}_{N} : \R^n \times \R^n \to \R$ and $\rho^{\cal S}_{1,N} : \R^n \to \R$ given by \eqref{eq:QN-spin-defi}, \eqref{eq:AutoSpin-gleich} and \eqref{eq:one-body-density-spin}, respectively.

\subsection{Semiclassical Weyl quantization}\label{sec:Weyl}

For easy reference we collect some background material on the Weyl pseudo\-differential calculus that will be needed in the sequel. We will restrict to the very basic symbol class 
$$ S 
   := S(\R^{2n}) 
   := \{ a \in C^{\infty}(\R^{2n}) : 
   \| \partial^{\alpha} a \|_{L^{\infty}(\R^{2n})} < \infty \mbox{ for all multiindices } \alpha \} $$ 
of smooth functions with bounded derivatives, which will be sufficient for our applications. For much more general classes and additional material on semiclassical analysis see, e.g., \cite{DimassiSjoestrand,EvansZworski,Martinez}. 

For $h > 0$ a (small) positive number, the semiclassical Weyl quantization of a symbol $a \in S$ is the operator $a^{\rm w}(x, hD) : {\cal S}(\R^n) \to {\cal S}(\R^n)$ acting on Schwartz functions defined by
$$ a^{\rm w}(x, hD) u(x) 
   := (2\pi)^{-n} \int_{\R^n} e^{i (x - y) \cdot \xi} \, 
   a \left( \frac{x + y}{2}, h \xi \right) u(y) \, dy \, d\xi. $$ 
$a^{\rm w}(x, hD)$ extends to a bounded operator from $L^2(\R^n)$ into itself and one has $(a^{\rm w}(x, hD))^* = \overline{a}^{\rm w}(x, hD)$. In particular, if $a$ is real valued, then $a^{\rm w}(x, hD)$ is self-adjoint. 

If $a,b \in S$ are two symbols, then also their product generates a pseudifferential operator and there exists a constant $C = C(a, b)$ (independent of $h$) such that 
\begin{align}\label{eq:compose}
  \| a^{\rm w}(x, hD) \, b^{\rm w}(x, hD) - 
     (ab)^{\rm w}(x, hD) \|_{L^2 \to L^2} 
  \le C h.  
\end{align}
G\r{a}rding's inequality asserts that there exists a constant $C = C(a)$ such that for all sufficiently small $h > 0$ and all $u \in L^2(\R^n)$ 
\begin{align}\label{eq:Garding}
  ( u, a^{\rm w}(x, hD) u )_{L^2} \ge - C h \| u \|_{L^2}^2, 
\end{align}
whenever $a \in S$ satisfies $a \ge 0$, i.e., is real valued and non-negative. 
  
Now if $a \in {\cal S}(\R^{2n})$ is itself a Schwartz function on phase space, then the associated Weyl quantization is given by 
\begin{align}\label{eq:a-trace}
  a^{\rm w}(x, hD) u(x) 
  = \int_{\R^n} k_{a,h}(x, y) \, u(y) \, dy 
\end{align}
for all $u \in L^2(\R^n)$, where the integral kernel $k_{a,h} \in {\cal S}(\R^{2n})$ is given by 
\begin{align}\label{eq:a-kernel} 
  k_{a,h}(x, y) 
  = (2\pi)^{-n} \int_{\R^n} e^{i (x - y) \cdot \xi} \, 
  a \left( \frac{x + y}{2}, h \xi \right) \, d\xi. 
\end{align}

\section{Localized spectral asymptotics}\label{sec:loc-spec}

In this section we will prove Theorem \ref{theo:conv-to-trace}. For the following preparatory lemmas we fix $a \in S$ of the form $a(x,\xi) = a_1(x) a_2(\xi)$, where $a_1 \in C^{\infty}_c(\Omega)$ and $a_2 \in C^{\infty}(\R^n)$ has bounded derivatives of all orders. 

\begin{lemma}\label{lemma:localize}
Let $0 < \alpha < \frac{1}{2}\min\{1, \gamma\}$, $m \in \N$. Suppose $\eta_1, \eta_2, \chi \in C^{\infty}(\R^n)$ are bounded and such that $\supp \eta_1 \subset B_{\gamma+\alpha}$,  $\supp \eta_2 \subset \R^n \setminus \overline{B}_{\gamma-\alpha}$ and $\supp \chi \subset \subset \Omega$. There exist constants $\beta$ (independent of $k$, $N$ and $\alpha$) and $C$ (independent of $k$ and $N$) such that for sufficiently large $N$: 
\begin{itemize}
\item[(i)] $\alpha \beta N \le k \le (1 - \alpha \beta)N$ implies 
$$ \| (1 \otimes \eta_j \, a \, 1 \otimes \eta_2)^{\rm w} (x, N^{-1/n}D) \chi u_k \| \le C N h_k^m, \quad j = 1,2 $$ 
\item[(ii)] and $k \ge (1 + \alpha \beta)N$ implies 
$$ \| (1 \otimes \eta_1 \, a \, 1 \otimes \eta_1)^{\rm w} (x, N^{-1/n}D) \chi u_k \| \le C N h_k^m. $$  
\end{itemize}
\end{lemma} 
With a slight abuse of notation we will sometimes simply write $\eta_i$ for the function $(x,\xi) \mapsto 1 \otimes \eta_i(x,\xi) = \eta_i(\xi)$. \smallskip 

\Proof If $N^{-1/n} \xi \in \supp \eta_1$, then by Weyl's law we have 
\begin{align*}
  |h_k \xi| 
  &= |(\gamma + o(1))^{-1} k^{-1/n} \xi| \\ 
  &\le |(\gamma + o(1))^{-1} (1 + \alpha \beta)^{-1/n} N^{-1/n} \xi| \\ 
  &\le (1 + \alpha) \gamma^{-1} (1 + \alpha \beta)^{-1/n} (\gamma + \alpha)
\end{align*}
for $k \ge (1 + \alpha \beta) N$, $N$ sufficiently large. For $\beta > 0$ suitably chosen, the right hand side of the last term is bounded by $1 - \alpha$ for any $0 < \alpha < \frac{1}{2}$, whence 
\begin{align}\label{eq:nennerabschi}
|h_k \xi| \le 1 -  \alpha. 
\end{align}
Analogously one obtains 
\begin{align}\label{eq:nennerabschii}
  |h_k \xi| 
  &\ge 1 + \alpha 
\end{align}
for $|N^{-1/n} \xi| \ge \gamma - \alpha$ and $\alpha \beta N \le k \le (1 - \alpha \beta) N$, $N$ sufficiently large. (The necessary size of $N$ may depend on $\alpha$ and $\beta$.) 

If $i=1$ and $k \ge (1 + \alpha \beta) N$ or if $i=2$ and $\alpha \beta N \le k \le (1 - \alpha \beta) N$, then 
\begin{align*}
  &(\eta_j a \eta_i)^{\rm w} (x, N^{-1/n} D) \chi u_k (x) \\ 
  &= (2\pi)^{-n} \int_{\R^{2n}} e^{i (x - y) \cdot \xi} \eta_j(N^{-1/n} \xi) \eta_i(N^{-1/n} \xi) 
     a\left(\frac{x + y}{2}, N^{-1/n} \xi \right) \chi(y) u_k(y) \, dy \, d\xi \\ 
  &= (2\pi)^{-n} \int_{\R^n} e^{i x \cdot \xi} f_{k,N,m,i,j}(\xi) 
     (|h_k \xi|^2 - 1)^m \int_{\R^n} e^{-i y \cdot \xi} g_{k}(x,y) \, dy \, d\xi 
\end{align*}
for each $m \in \N$, where $f_{k,N,m,i,j}(\xi) = \frac{\eta_j(N^{-1/n} \xi) \eta_i(N^{-1/n} \xi)}{(|h_k \xi|^2 - 1)^m} a_2(N^{-1/n} \xi)$ and $g_{k}(x,y) = a_1(\frac{x + y}{2}) \chi(y) u_k(y)$. 

It follows that $(\eta_j a \eta_i)^{\rm w} (x, N^{-1/n} D) \chi u_k (x) = 0$ for $x \notin 2 \Omega - \Omega$. For general $x$, $|(\eta_j a \eta_i)^{\rm w} (x, N^{-1/n} D) \chi u_k (x)|$ is bounded by 
\begin{align*}
  &(2\pi)^{-n} \| f_{k,N,m,i,j} \|_{L^2} 
  \| {\cal F}_y (- h_k^2 \Delta - 1)^m g_{k}(x, \cdot) \|_{L^2} \\ 
  &= (2\pi)^{-n/2} \| f_{k,N,m,i,j} \|_{L^2} 
  \| (- h_k^2 \Delta - 1)^m g_{k}(x, \cdot) \|_{L^2}, 
\end{align*}
where ${\cal F}_y$ denotes Fourier transform with respect to the $y$-variable only: $({\cal F}_y v(x, \cdot))(\eta) = \int_{\R^n} e^{-i \eta \cdot y} v(x, y) \, dy$. Since by \eqref{eq:nennerabschi} and \eqref{eq:nennerabschii}
\begin{align*}
  \| f_{k,N,m,i,j} \|_{L^2}^2 
  &\le \sup_{\xi} \left| \frac{\eta_j(N^{-1/n} \xi) \eta_i(N^{-1/n} \xi)}{(|h_k \xi|^2 - 1)^m} \right|^2 
  \int_{\R^n} \left| a_2(N^{-1/n} \xi) \right|^2 \, d\xi \\ 
  &\le \alpha^{-2m} \| \eta_j \eta_i\|_{L^{\infty}} N \| a_2 \|_{L^2}^2  
\end{align*}
and moreover Lemma \ref{lemma:int-reg-absch} below shows that there exists a constant $C$ only depending on $m$, $a_1$ and $\chi$ such that 
$$ \| (- h_k^2 \Delta - 1)^m g_{k}(x, \cdot) \|_{L^2} 
   \le C h_k^m $$
for all $k$, it thus follows that 
\begin{align*} 
  |(\eta_j a \eta_i)^{\rm w} (x, N^{-1/n} D) \chi u_k (x)|
  \le C(\alpha) N h_k^m. 
\end{align*}
\eop

\begin{lemma}\label{lemma:int-reg-absch}
Let $\psi(y) = \psi_x(y) = a_1(\frac{x+y}{2}) \chi(y)$. There exist constants $C = C(m,\psi)$ such that for all $k \in \N$ and $x \in \R^n$ 
$$ \| (- h_k^2 \Delta - 1)^m \psi u_k \|_{L^2} 
   \le C h_k^m. $$
\end{lemma} 

\Proof By induction on $m$ it follows that 
$$ (- h_k^2 \Delta - 1)^m \psi u_k 
   = \sum_{|\alpha| \le m} h_k^{2m} \, g_{m,\alpha} \, \partial^{\alpha} u_k $$ 
for suitable smooth functions $g_{m,\alpha}$ supported on $\supp \psi$. But then 
\begin{align*}
  \| (- h_k^2 \Delta - 1)^m \psi u_k \|_{L^2(\R^n)} 
  \le C\sum_{|\alpha| \le m} h_k^{2m} \| \partial^{\alpha} u_k \|_{L^2(\supp \psi)} 
\end{align*} 
with $C$ only depending on $a_1$ and $\chi$. 

Consider a nested sequence $\Omega \supset \supset \Omega_0 \supset \supset \Omega_1 \supset \supset \ldots \supset \supset \supp \psi$ and first observe that $\| u_k \|_{L^2(\Omega_0)}, h_k \| \nabla u_k \|_{L^2(\Omega_1)} \le C$. The first inequality is clear. For the second estimate consider a cut-off function $\theta \in C^{\infty}_c(\Omega)$ with $0 \le \theta \le 1$ and $\theta \equiv 1$ on $\Omega_1$. Then 
\begin{align*}
  \| \theta \nabla u_k \|_{L^2(\Omega)}^2 
  &= \int_{\Omega} \theta^2 |\nabla u_k|^2   
  = - \int_{\Omega} \theta^2 \, u_k \, \Delta u_k - 2 \int_{\Omega} \theta \, u_k \, \nabla \theta \cdot \nabla u_k \\ 
  &\le h_k^{-2} + 2 \| \theta \nabla u_k \|_{L^2(\Omega)} \| u_k \nabla \theta \|_{L^2(\Omega)} 
  \le h_k^{-2} + C \| \theta \nabla u_k \|_{L^2(\Omega)}  
\end{align*}
and so $\| \nabla u_k \|_{L^2(\Omega_1)} \le \| \theta \nabla u_k \|_{L^2(\Omega)} \le C h_k^{-1}$. 

Standard interior regularity estimates (see, e.g., \cite{GilbargTrudinger}) for $\Delta \partial^{\alpha} u_k = h_k^{-2} \partial^{\alpha} u_k$ on $\Omega_{|\alpha|}$ give 
\begin{align*}
  h_k^{|\alpha| + 2} \| \partial_{ij} \partial^{\alpha} u_k \|_{L^2(\Omega_{|\alpha|+2})} 
  &\le C h_k^{|\alpha| + 2} \| \partial^{\alpha} u_k \|_{H^2(\Omega_{|\alpha|+2})} \\ 
  &\le C h_k^{|\alpha| + 2} \left( \| \partial^{\alpha} u_k \|_{L^2(\Omega_{|\alpha|})} 
       + \| h_k^{-2} \partial^{\alpha} u_k \|_{L^2(\Omega_{|\alpha|})} \right),
\end{align*}
and so $h_k^{|\alpha|} \| \partial^{\alpha} u_k \|_{L^2(\Omega_{|\alpha|})} \le C$ by induction. (Alternatively one could of course directly refer to regularity estimates for powers of $\Delta$.) Inserting this estimate above, we find 
\begin{align*}
  \| (- h_k^2 \Delta - 1)^m \psi u_k \|_{L^2} 
  \le C\sum_{|\alpha| \le m} h_k^{2m} h_k^{-|\alpha|} 
  \le C h_k^m.  
\end{align*} 
\eop

If $0 < \eps < \frac{1}{2}$ is small enough, we may choose smooth and bounded cut-off functions $0 \le \eta_1, \eta_2, \chi_1 \le 1$ on $\R^n$ such that 
$$ \supp \eta_1 \subset B_{\gamma+\eps},~
   \supp \eta_2 \subset \R^n \setminus \overline{B}_{\gamma-\eps} \mbox{ and } 
   \supp \chi_1 \subset \{x \in \Omega : \dist(x, \partial\Omega) > \eps \} $$ 
which satisfy 
$$ \eta_1 + \eta_2  \equiv 1 \mbox{ on } \R^n 
   \mbox{ and } 
   \chi \equiv 1 \mbox{ on } \supp a_1. $$ 
Viewing $\chi$ as a multiplication operator we can state the following lemma.
\begin{lemma}\label{lemma:smoothing} 
There exist constants $c$ (independent of $\eps$ and $N$) and $C(\eps)$ (independent of $N$) such that 
\begin{align*}
  &\Bigg| \sum_{k = 1}^N ( u_k, a^{\rm w}(x, N^{-1/n}D) u_k ) 
  - \trace ( \chi (\eta_1 a \eta_1)^{\rm w}(x, N^{-1/n}D) \chi ) \Bigg| \\ 
  &\le c \eps N + C(\eps) N^{1-1/n}. 
\end{align*}
\end{lemma}

\Proof By \eqref{eq:compose}, 
\begin{align}\label{eq:a-localize}
\| a^{\rm w}(x, N^{-1/n} D) - \chi a^{\rm w}(x, N^{-1/n} D) \chi \|_{L^2 \to L^2} \le C(\eps) N^{-1/n} 
\end{align} 
for some constant $C(\eps)$. Noting that 
$$ a^{\rm w} 
   = (\eta_1 a \eta_1)^{\rm w} + 2 (\eta_1 a \eta_2)^{\rm w} + (\eta_2 a \eta_2)^{\rm w} $$
(where here and in the sequel we sometimes drop the argument $(x, N^{-1/n}D)$ of the Weyl quantized operators) and choosing $\beta$ according to Lemma \ref{lemma:localize} with $\alpha = \eps$ and $m = 3n$, we obtain 
\begin{align}\label{eq:kleine-k}
  | ( u_k, \chi a^{\rm w} \chi u_k ) - (u_k, \chi (\eta_1 a \eta_1)^{\rm w} \chi u_k) | 
  \le C(\eps) N h_k^{3n} 
  \le C(\eps) N^{-1} 
\end{align} 
for $\eps \beta N \le k \le (1 - \eps \beta) N$, $N$ sufficiently large. (Note that $h_k^{3n} \le 2 \gamma^{-3n} k^{-3}$ for $k \ge \eps \beta N$ if $N$ is large enough.) On the other hand, 
\begin{align}\label{eq:grosse-k}
  | ( u_k, \chi (\eta_1 a \eta_1)^{\rm w} \chi u_k) | 
  \le C(\eps) N h_k^{3n}  
\end{align} 
for $k \ge (1 + \eps \beta) N$, $N$ sufficiently large, again by Lemma \ref{lemma:localize}. 

Now $a^{\rm w}$, $(\eta_1 a \eta_1)^{\rm w}$ and multiplication by $\chi$ are bounded operators on $L^2$ with operator norm bounded independently of $N$. From \eqref{eq:a-localize}, \eqref{eq:kleine-k} and \eqref{eq:grosse-k} we therefore obtain 
\begin{align*}
  \Bigg| \sum_{k = 1}^N ( u_k, a^{\rm w} u_k ) 
   - \sum_{k = 1}^{\infty} 
     ( u_k, \chi (\eta_1 a \eta_1)^{\rm w} \chi u_k ) \Bigg|   
  \le C(\eps) N^{1-1/n} + c \eps N + C(\eps) N \sum_{k = N + 1}^{\infty} h_k^{3n} 
\end{align*}
with $c$ not depending on $\eps$. The claim now follows from $h_k^{3n} \le 2 \gamma^{-3n} k^{-3}$ for large $k$ and thus $\sum_{k = N + 1}^{\infty} h_k^{-3n} \le \gamma^{-3n} N^{-2}$ for large $N$ and the fact that $\chi (\eta_1 a \eta_1)^{\rm w} \chi$ vanishes on functions supported in $\R^n \setminus \Omega$. \eop

Since $\eta_1 a \eta_1 \in C^{\infty}_c(\R^{2n})$, $(\eta_1 a \eta_1)^{\rm w}(x, N^{-1/n}D)$ is of trace class and thus also $A_N = \chi (\eta_1 a \eta_1)^{\rm w} \chi$ is of trace class with  
$$ A_N v = \int_{\R^n} k_N(x, y) v(y) \, dy \quad \forall \, v \in {\cal S}(\R^n), $$ 
where $k_N$ is the integral kernel 
\begin{align*}
  & k_N(x, y) 
= (2\pi)^{-n} \int_{\R^{n}} e^{i (x - y) \cdot \xi} \eta_1^2(N^{-1/n} \xi) \chi(x) \chi(y)  
     a\left(\frac{x + y}{2}, N^{-1/n} \xi \right) \, d\xi  
\end{align*}
(cf.\ \eqref{eq:a-kernel}). As $k_N \in C^{\infty}_c(\R^{2n})$, the trace of $A_N$ can be computed by integrating the diagonal elements of the kernel. (This is in fact true for more general kernels, cf.\ \cite{Brislawn}.) Hence, 
\begin{align}\label{eq:trace-A}
  \trace A_N 
  &= \int_{\R^n} k_N(x, x) \, dx \nonumber \\ 
  &= (2\pi)^{-n} \int_{\R^{2n}} \eta_1^2(N^{-1/n} \xi) \chi^2(x)  
     a(x, N^{-1/n} \xi) \, d\xi \, dx \nonumber \\ 
  &= (2\pi)^{-n} N \int_{\R^{2n}} \eta_1^2(\xi) \chi^2(x)  
     a(x, \xi) \, d\xi \, dx. 
\end{align}

For easy reference we summarize our observations so far in the following lemma. 
\begin{lemma}\label{lemma:conv-to-trace} 
For $a = a_1 \otimes a_2 \in S$ with $a_1 \in C^{\infty}_c(\Omega)$ one has 
$$ \lim_{N \to \infty} N^{-1} \sum_{k = 1}^N ( u_k, a^{\rm w}(x, N^{-1/n}D) u_k ) 
   = (2\pi)^{-n} \int_{\Omega \times B_{\gamma}} a(x, \xi) \, dx \, d\xi. $$ 
\end{lemma}

\Proof Lemma \ref{lemma:smoothing} and Equation \eqref{eq:trace-A} imply that 
\begin{align*}
  &\limsup_{N \to \infty} \Bigg| N^{-1} \sum_{k = 1}^N ( u_k, a^{\rm w}(x, N^{-1/n}D) u_k ) \\ 
  &\qquad \qquad \qquad 
  - (2\pi)^{-n} \int_{\R^{2n}} \chi^2(x) \eta_1^2(\xi) a(x, \xi) \, dx \, d\xi\Bigg|
  \le c \eps. 
\end{align*}
By our choice of $\chi$ and $\eta_1$ we thus have 
\begin{align*}
  \limsup_{N \to \infty} \Bigg| N^{-1} \sum_{k = 1}^N ( u_k, a^{\rm w}(x, N^{-1/n}D) u_k ) 
  - (2\pi)^{-n} \int_{\Omega \times B_{\gamma}} a(x, \xi) \, dx \, d\xi \Bigg|
  \le c \eps 
\end{align*}
with $c$ independent of $\eps$. The claim now follows since $\eps $ was arbitrary. \eop

In order to prove Theorem \ref{theo:conv-to-trace} it remains to show that the restrictive assumption on the form of $a$ in Lemma \ref{lemma:conv-to-trace} can be dropped. \smallskip 

\noindent {\em Proof of Theorem \ref{theo:conv-to-trace}.} Let $a \in S$ and suppose first that $a(x,\xi)$ is real-valued, non-negative and vanishes for $x \notin \Omega_{\rho} = \{z \in \Omega : \dist(z, \partial \Omega) \ge \rho\}$ for some $\rho > 0$. For given $\eps > 0$ approximating $a$ uniformly by symbols of the form $a_{\eps}(x, \xi) = \sum_{i,j \in \Z^n} a(\delta i,\delta j) \psi_i^{\delta}(x) \psi_j^{\delta}(\xi)$, where $(\psi_i^{\delta})$ is a partition of unity on $\R^n$ with $\supp \psi_i^{\delta} \subset \delta i + (0, 2\delta)^n$, $\delta$ sufficiently small, by G\r{a}rding's inequality \eqref{eq:Garding} we find 
$$ ( u_k, \pm (a - a_{\eps})^{\rm w} (x, N^{-1/n}D) u_k )_{L^2} 
   \ge - C N^{-1/n} - \eps. $$ 
As a consequence, the assertion of Lemma \ref{lemma:conv-to-trace} remains true for the symbol $a$. 

Now more generally consider $a \in S$, $a \ge 0$ and let $\rho > 0$. Choose a cut-off function $\theta \in C^{\infty}_c(\Omega)$ with $0 \le \theta \le 1$ and $\theta \equiv 1$ on $\Omega_{\rho}$. Let $a' = (1 - \theta \otimes 1) a$ and suppose first that $a \ge 0$. By G\r{a}rding's inequality \eqref{eq:Garding} we obtain 
$$ ( u_k, (a')^{\rm w} (x, N^{-1/n}D) u_k )_{L^2} 
   \ge - C N^{-1/n} $$ 
and therefore 
\begin{align}\label{eq:unten-aussen}
  \liminf_{N \to \infty} N^{-1} \sum_{k = 1}^N ( u_k, (a')^{\rm w}(x, N^{-1/n}D) u_k )_{L^2} 
  \ge 0. 
\end{align}

Now setting $\tilde{a} = m \theta \otimes 1 \in S$ for $m = \| a \|_{\infty}$, again by G\r{a}rding's inequality \eqref{eq:Garding} we obtain 
$$ ( u_k, (m - a' - \tilde{a})^{\rm w} (x, N^{-1/n}D) u_k )_{L^2} 
   \ge - C N^{-1/n}. $$ 
From Lemma \ref{lemma:conv-to-trace} and Equation \eqref{eq:W-Mass} we can now infer that 
\begin{align}\label{eq:oben-aussen}
  &\limsup_{N \to \infty} N^{-1} \sum_{k = 1}^N ( u_k, (a')^{\rm w} (x, N^{-1/n}D) u_k )_{L^2} \nonumber \\ 
  &\le \limsup_{N \to \infty} N^{-1} \sum_{k = 1}^N 
   \left( m - ( u_k, (\tilde{a})^{\rm w} (x, N^{-1/n}D) u_k )_{L^2} \right) \nonumber \\ 
  &= m - (2\pi)^{-n} \int_{\Omega \times B_{\gamma}} \tilde{a}(x, \xi) \, dx \, d\xi \nonumber \\ 
  &\le (2\pi)^{-n} m |(\Omega \setminus \Omega_{\rho}) \times B_{\gamma}|, 
\end{align}
which tends to zero as $\rho \to 0$. 

On the other hand Lemma \ref{lemma:conv-to-trace} yields 
\begin{align}\label{eq:innen}
  \lim_{N \to \infty} N^{-1} \sum_{k = 1}^N ( u_k, (a'')^{\rm w} (x, N^{-1/n}D) u_k )_{L^2} 
  = (2\pi)^{-n} \int_{\Omega \times B_{\gamma}} a''(x, \xi) \, dx \, d\xi, 
\end{align}
where $a'' = (\theta \otimes 1) \, a$, which converges to $(2\pi)^{-n} \int_{\Omega \times B_{\gamma}} a(x, \xi) \, dx \, d\xi$ as $\rho \to 0$. Summarizing \eqref{eq:unten-aussen}, \eqref{eq:oben-aussen} and \eqref{eq:innen} we see that indeed $a = a' + a''$ satisfies 
\begin{align}\label{eq:to-show}
  \lim_{N \to \infty} N^{-1} \sum_{k = 1}^N ( a^{\rm w} (x, N^{-1/n}D) u_k, u_k )_{L^2} 
  = (2\pi)^{-n} \int_{\Omega \times B_{\gamma}} a(x, \xi) \, dx \, d\xi. 
\end{align}

For general $a \in S$ we may choose $\mu \in \R$ such that $\Re a + \mu, \Im a + \mu \ge 0$. With the help of \eqref{eq:W-Mass} the above calculations applied to $\Re a + \mu$ and $\Im a + \mu$ give \eqref{eq:to-show}. \eop

\section{Averaged correlation asymptotics of eigenfunctions}\label{sec:averaged-ev}

We wish to apply the results of the previous section to investigate convergence properties of the pair correlation function \eqref{eq:AutoSpin-gleich}. In order to do so, we will investigate the unsquared one-body density matrix expression from \eqref{eq:QN-defi}: 
$$ Q_N(x, y) := N^{-1} \sum_{k = 1}^N 
  u_k \left( x + \frac{N^{-1/n} y}{2} \right) \, \overline{u_k}\left( x - \frac{N^{-1/n} y}{2} \right). $$
\begin{proposition}\label{prop:QN-toQ}
$Q_N$ converges in distributions to $Q$ with 
$$ Q(x, y) 
   = \frac{2^{n/2} \Gamma(\frac{n}{2} + 1) J_{n/2}(\gamma |y|)}{|\Omega|(\gamma |y|)^{n/2}} \chi_{\Omega}(x). $$ 
\end{proposition}

\Proof With the help of \eqref{eq:a-trace} and \eqref{eq:a-kernel} it is not hard to see that  the formula 
$$ \langle T_{k,N}, a \rangle 
   = (u_k, a^{\rm w}(x, N^{-1/n} D) u_k) \quad 
   \forall\, a \in {\cal S}(\R^n \times \R^n). $$ 
defines a tempered distribution $T_{k,N} \in {\cal S}'(\R^n \times \R^n)$. Since for $a \in {\cal S}(\R^n \times \R^n)$ 
\begin{align*}
  &\langle T_{k,N}, a \rangle 
  = \int_{\R^n} (2\pi)^{-n} \int_{\R^{2n}} e^{i (x - y) \cdot \xi} 
     a \left( \frac{x + y}{2}, N^{-1/n} \xi \right) u_k(y) \, dy \, d\xi \, \overline{u_k}(x) \, dx \\ 
  &= (2 \pi)^{-n} \int_{\R^{3n}} e^{-i w \cdot \zeta} a(z, \zeta) 
     u_k \left( z + \frac{N^{-1/n} w}{2} \right) 
     \overline{u_k} \left( z - \frac{N^{-1/n} w}{2} \right) \, dw \, dz \, d\zeta, 
\end{align*}
where we have changed coordinates according to $z = \frac{x + y}{2}$, $w = N^{1/n} (y - x)$ and $\zeta = N^{-1/n} \xi$, $T_{k,N}$ is the $L^{\infty}$-function 
$$ T_{k,N}(z,\zeta) 
   = (2 \pi)^{-n} {\cal F} \left[ u_k \left( z + \frac{N^{-1/n} (\cdot)}{2} \right) 
     \overline{u_k} \left( z - \frac{N^{-1/n} (\cdot)}{2} \right) \right] (\zeta). $$ 
Denoting by ${\cal F}_y Q(x,y)$ the partial Fourier transform with respect to the second variable only, we find 
\begin{align*}
  U_N(x,\xi) 
  := (2 \pi)^{-n} {\cal F}_{y} Q_N(x, \xi)
  = N^{-1} \sum_{k = 1}^N T_{k,N}(x,\xi). 
\end{align*}

Now observe that Theorem \ref{theo:conv-to-trace} implies that, as $N \to \infty$,
\begin{align}\label{eq:UN-conv}
  \langle U_N, a \rangle 
  \to (2 \pi)^{-n} \int_{\Omega \times B_{\gamma}} a(x, \xi) \, dx \, d\xi 
\end{align}
and so 
\begin{align*}
  &\int_{\R^{2n}} Q_N(x, y) \, a(x, y) \, dx \, dy 
   = (2 \pi)^n \int_{\R^{2n}} {\cal F}_{y}^{-1} U_N(x,y) \, a(x, y) \, dx \, dy \\ 
  &= (2 \pi)^n \int_{\R^{2n}} U_N(x,y) {\cal F}_y^{-1} \, a(x, y) \, dx \, dy 
   \to \int_{\R^{2n}} \chi_{\Omega \times B_{\gamma}} (x, y) \, {\cal F}_y^{-1} a(x, y) \, dx \, dy \\
  &= \int_{\R^{2n}} \chi_{\Omega}(x) {\cal F}^{-1} \chi_{B_{\gamma}} (y) \, a(x, y) \, dx \, dy 
   = (2 \pi)^{-n} \int_{\R^{2n}} \chi_{\Omega}(x) \widehat{\chi_{B_{\gamma}}}(y) \, a(x, y) \, dx \, dy,
\end{align*}
which shows that $Q_{N}$ converges in distributions to 
\begin{align}\label{eq:Q-formulae}
  Q(x, y) 
   = \frac{\chi_{\Omega}(x) \widehat{\chi_{B_{\gamma}}}(y)}{(2 \pi)^n} 
   = \frac{\chi_{\Omega}(x) \widehat{\chi_{B_{\gamma}}}(y)}{|\Omega \times B_{\gamma}|}
   = \frac{2^{n/2} \Gamma(\frac{n}{2} + 1) J_{n/2}(\gamma |y|)}{|\Omega|(\gamma |y|)^{n/2}} \chi_{\Omega}(x), 
\end{align}
because $\gamma^n = \frac{(2\pi)^n}{|\Omega \times B_{1}|}$ and $\widehat{\chi_{B_1}}(y) = (2 \pi)^{n/2} |y|^{-n/2} J_{n/2}(|y|)$ and $\chi_{B_{\gamma}} = \chi_{B_1}(\cdot/\gamma)$. \eop

The second auxiliary result for the proof of Theorem \ref{theo:one-body-matrix} is the following. 
\begin{proposition}\label{prop:QN-bounded}
$Q_N$ is bounded uniformly on compact subsets of $\Omega \times \R^n$.
\end{proposition}

In order to prove this result it will be convenient to consider the function 
\begin{align}\label{eq:QN-tilde-defi}
  \tilde{Q}_N(x, y) :=  Q_N \left( x + \frac{N^{-1/n} y}{2}, y \right). 
\end{align}
\begin{lemma}\label{lemma:higher-reg}
Let $\Omega' \subset \subset \Omega$, $R > 0$. There exists a constant $C = C(R)$ such that for all $N$ sufficiently large
$$ \sup_{y \in B_R} | \tilde{Q}_N(x, y) |^2  
   \le C N^{-1} \sum_{k = 1}^N |u_k(x)|^2 $$
for all $x \in \Omega'$. 
\end{lemma}

\Proof If $x \in \Omega'$ and $N$ is large enough, then $x + 2N^{-1/n}B_R \subset \Omega$, so that $y \mapsto \tilde{Q}_N(x, y) = N^{-1} \sum_{k = 1}^N u_k(x + N^{-1/n}y) \overline{u_k}(x)$ is smooth on $B_{2R}$ and we may calculate 
\begin{align*}
  \Delta_y^m \tilde{Q}_N(x, y) 
  &= \Delta_y^m N^{-1} \sum_{k = 1}^N u_k(x + N^{-1/n} y) \, \overline{u_k}(x) \\ 
  &=  N^{-1} \sum_{k = 1}^N \lambda_k^m N^{-2m/n} u_k(x + N^{-1/n} y) \, \overline{u_k}(x). 
\end{align*}
Weyl's law guarantees that $\lambda_k \le C k^{2/n}$. With $z = x + N^{-1/n}y$ we thus find 
\begin{align*}
  \int_{B_{2R}} \left| \Delta_y^m \tilde{Q}_N(x, y) \right|^2 \, dy 
  &\le N^{-2} \int_{\R^n} \left| \sum_{k = 1}^N \lambda_k^m N^{-2m/n} 
     u_k(x + N^{-1/n} y) \, \overline{u_k}(x) \right|^2 \, dy \\ 
  &= N^{-1} \int_{\R^n} \left| \sum_{k = 1}^N \lambda_k^m N^{-2m/n} 
     \, \overline{u_k}(x) \, u_k(z) \right|^2 \, dz \\ 
  &= N^{-1} \sum_{k=1}^N \lambda_k^{2m} N^{-4m/n} \, |u_k(x)|^2 \\ 
  &\le C(m) N^{-1} \sum_{k = 1}^N |u_k(x)|^2. 
\end{align*}
The claim now follows from interior regularity estimates on $B_R \subset B_{2R}$ for the Laplacian and Sobolev embedding. \eop

\par\noindent{\em Proof of Proposition \ref{prop:QN-bounded}.} 
Let $\Omega' \subset \subset \Omega$, $R > 0$. By Lemma \ref{lemma:higher-reg}
$$ | \tilde{Q}_N(x, 0) |^2 
   \le C N^{-1} \sum_{k = 1}^N |u_k(x)|^2 
   = C | \tilde{Q}_N(x, 0) | $$
for all $x \in \Omega'$ and so $|\tilde{Q}_N(x, 0)| \le C$. Now using Lemma \ref{lemma:higher-reg} again we find 
$$ \sup_{y \in B_R} | \tilde{Q}_N(x, y) |^2  
   \le C N^{-1} \sum_{k = 1}^N |u_k(x)|^2 
   = C | \tilde{Q}_N(x, 0) | 
   \le C. $$
Hence $\tilde{Q}_N$ and thus also $Q_N$ is bounded uniformly on compact subsets of $\Omega \times \R^n$. \eop

\noindent {\em Proof of Theorem \ref{theo:one-body-matrix}.} 
By Proposition \ref{prop:QN-bounded} it remains to show that the convergence $Q_N \to Q$ in Proposition \ref{prop:QN-toQ} is in fact strong in $L^2$. To this end, by Proposition \ref{prop:QN-toQ} it is enough to prove that $\| Q_N \|_{L^2} \to \| Q \|_{L^2}$. 

Abbreviating $y_N := \frac{N^{-1/n} y}{2}$ and changing coordinates according to $w = x + y_N$, $z = x - y_N$, 
so that $|\det \nabla_{(w,z)} (x,y)| = N$, we find 
\begin{align*}
  &\int_{\R^{2n}} | Q_N(x, y) |^2 \, dx \, dy \\ 
  &= N^{-2} \sum_{k,m = 1}^N 
  \int_{\R^{2n}} u_k \left( x + y_N \right) \, \overline{u_k}\left( x -y_N \right) \, 
  \overline{u_m} \left( x + y_N \right) \, u_m\left( x - y_N \right)  \, dx \, dy \\ 
  &=  N^{-1} \sum_{k,m = 1}^N 
  \int_{\R^{2n}} u_k(w) \, \overline{u_m} (w) \, 
  \overline{u_k} (z) \, u_m(z)  \, dw \, dz 
  =  N^{-1} \sum_{k,m = 1}^N \delta_{km}^2 
  = 1. 
\end{align*}
The claim now follows since also $Q(x,y) = \frac{\chi_{\Omega}(x) \widehat{\chi_{B_{\gamma}}}(y)}{(2 \pi)^n}$ (cf.\ \eqref{eq:Q-formulae}) satisfies 
\begin{align*} 
  \int_{\R^{2n}} | Q(x, y) |^2 \, dx \, dy
  = \frac{|\Omega|}{(2\pi)^{2n}} \int_{\R^n} \left| \widehat{\chi_{B_{\gamma}}}(y) \right|^2 \, dy 
  = \frac{|\Omega|}{(2\pi)^{n}} \int_{\R^n} \left| \chi_{B_{\gamma}}(y) \right|^2 \, dy 
  =  1 
\end{align*}
because $\frac{|\Omega| |B_{\gamma}|}{(2\pi)^{n}} = \gamma^n \frac{|\Omega \times B_1|}{(2\pi)^{n}} = 1$. 
\eop 

\par \noindent {\em Proof of Theorem \ref{theo:wigner-seitz}.} 
This is immediate from Theorem \ref{theo:one-body-matrix}. \eop 

\par\noindent{\em Proof of Theorem \ref{theo:one-body-conv}.} 
With \eqref{eq:one-body-density}, \eqref{eq:QN-defi} and \eqref{eq:QN-tilde-defi} we can write  
$$ \rho_{1, N}(x) 
   = N^{-1} \sum_{k = 1}^N u_k(x) \overline{u_k}(x) 
   = Q_N(x, 0) 
   = \tilde{Q}_N(x, 0).$$ 
Thanks to Proposition \ref{prop:QN-bounded} it suffices to establish the $L^1$-convergence of $\rho_{1, N}$ to $\bar{\rho}$. By Theorem \ref{theo:one-body-matrix} $Q_{N}$, and thus also $\tilde{Q}_N$, converges to $Q$ in $L^2(\R^{2n})$, where $Q(x,y) = \frac{\widehat{\chi_{B_{\gamma}}}(y)\chi_{\Omega}(x)}{|\Omega \times B_{\gamma}|}$, and so $\int_{\R^n} |\tilde{Q}_N(\cdot, y)|^2 \, dy \to \int_{\R^n} |\tilde{Q}(\cdot, y)|^2 \, dy$ in $L^1(\R^n)$. But -- as shown in the proof of Lemma \ref{lemma:higher-reg} -- 
\begin{align*}
  \int_{\R^n} \left| \tilde{Q}_N(x, y) \right|^2 \, dy
  = \rho_{1,N}(x)
\end{align*}
and also 
\begin{align*}
  \int_{\R^n} \left| Q(x, y) \right|^2 \, dy
  = \frac{\bar{\rho}(x) \|\widehat{\chi_{B_{\gamma}}}\|_{L^2(\R^n)}^2}{|\Omega| \, |B_{\gamma}|^2} 
  = \frac{(2\pi)^n \bar{\rho}(x)}{|\Omega| \, |B_{\gamma}|} 
  = \bar{\rho}(x), 
\end{align*}
which concludes the proof. \eop

\section{Correlation effects for the free Fermi gas}\label{sec:Fermi-gas-corr}

In this final section we will prove our correlation results for the free Fermi gas with spin: Theorems \ref{theo:one-body-matrix-spin}, \ref{theo:wigner-seitz-spin}, \ref{theo:one-body-conv-spin} and \ref{theo:Ex-LDA}. Here additional ``open shell effects'' have to be considered, which in part can be analyzed similarly as in the case of a three-dimensional box $(0, L)^3$, cf.\ \cite{Friesecke}. \smallskip 

\par\noindent{\em Proof of Theorem \ref{theo:one-body-matrix-spin}.} Let $a_N$ be the largest integer not exceeding $\frac{N}{m}$ such that $\lambda_{a_N+1} > \lambda_{a_N}$. Similarly, define $\tilde{a}_N$ to be the smallest integer which is not smaller than $\frac{N}{m}$ and satisfies $\lambda_{\tilde{a}_N+1} > \lambda_{\tilde{a}_N}$ By Weyl's law, $N = a_N m + b_N = \tilde{a}_N m - \tilde{b}_N$ with $0 \le b_N, \tilde{b}_N \ll N$. Then $\psi_{\alpha m + \beta} \mapsto u_{\alpha+1} \otimes \delta_{\beta}$ for $\alpha = 0, 1, \ldots, a_N - 1$, $\beta = 1,2,\ldots m$ extends to a unitary mapping of $\operatorname{span}\{ \psi_i : i \le a_N m \}$ onto itself. For $x', x'' \in \Omega$, $s_1,s_2 \in {\cal S}$ it follows that 
\begin{align}\label{eq:restterm}
  \sum_{i=1}^N \psi_{i}(x', s_1) \, \overline{\psi_{i}}(x'', s_2) 
  &= \sum_{\alpha=0}^{a_N-1} \sum_{\beta=1}^m 
    u_{\alpha+1}(x') \delta_{\beta}(s_1) \, \overline{u_{\alpha+1}}(x'') \delta_{\beta}(s_2) 
    + R_N \nonumber \\ 
  &= \sum_{\alpha=1}^{a_N} 
    u_{\alpha}(x') \, \overline{u_{\alpha}}(x'') \, \delta_{s_1 s_2} 
    + R_N, 
\end{align}
where $R_N = R_N(x',x'',s_1, s_2) = \sum_{i=a_Nm+1}^N \psi_{i}(x', s_1) \overline{\psi_{i}}(x'', s_2)$. 

The $L^2$-norm of the rest term can be estimated as follows. Setting $y_N = \frac{N^{-1/n} y}{2}$ and changing variables to $w = x + y_N$, $z = x - y_N$ so that $|\det \nabla_{(w,z)} (x,y)| = N$, we have
\begin{align*}
  &N^{-2} \sum_{s \in {\cal S}^2} \int_{\R^{2n}} | R_N(x + y_N, x - y_N, s_1, s_2) |^2 \, dx \, dy \\ 
  &= N^{-2} \sum_{s \in {\cal S}^2} \int_{\R^{2n}} \sum_{i,j = a_N m +1}^N 
     \psi_i \left( x + y_N, s_1 \right) \, \overline{\psi_i}\left( x - y_N, s_2 \right) \\ 
  &\hspace{4cm} \overline{\psi_j} \left( x + y_N, s_1 \right) \, \psi_j \left( x - y_N, s_2 \right)  \, dx \, dy \\ 
  &=  N^{-1} \sum_{i,j = a_N m +1}^N \sum_{s \in {\cal S}^2} 
  \int_{\R^{2n}} \psi_i \left( w, s_1 \right) \, \overline{\psi_j} \left( w, s_1 \right) \, 
     \overline{\psi_i}\left( z, s_2 \right) \, \psi_j \left( z, s_2 \right)  \, dw \, dz \\ 
  &=  N^{-1} \sum_{i,j = a_N m +1}^N  \delta_{ij}^2 
     = N^{-1} b_N 
\end{align*}
and hence 
\begin{align}\label{eq:rest-L2}
  N^{-1} \| R_N(x + y_N, x - y_N, s_1, s_2) \|_{L^2(\R^{2n})} \to 0 
\end{align} 
for $s_1, s_2 \in {\cal S}$. 

Thus, by \eqref{eq:restterm}, \eqref{eq:rest-L2} and Theorem \ref{theo:one-body-matrix}, we obtain 
\begin{align*} 
  &\lim_{N \to \infty} N^{-1} \sum_{i=1}^N \psi_{i}(x + y_N, s_1) \, \overline{\psi_{i}}(x - y_N, s_2) \\ 
  &= \lim_{N \to \infty} N^{-1} \sum_{\alpha=1}^{a_N} 
    u_{\alpha}(x + y_N) \overline{u_{\alpha}}(x - y_N) \delta_{s_1 s_2} \\ 
  &= \lim_{N \to \infty} \frac{a_N}{N} a_N^{-1} \sum_{\alpha=1}^{a_N} 
    u_{\alpha}\left( x + \left( \frac{a_N}{N} \right)^{1/n} \frac{a_N^{-1/n} y}{2} \right) 
    \overline{u_{\alpha}}\left( x - \left( \frac{a_N}{N} \right)^{1/n} \frac{a_N^{-1/n} y}{2} \right) \delta_{s_1 s_2} \\ 
  &= \frac{2^{n/2} \Gamma(\frac{n}{2} + 1) J_{n/2}(m^{-1/n} \gamma |y|) \delta_{s_1 s_2}}
     {m |\Omega|(m^{-1/n} \gamma |y|)^{n/2}} \chi_{\Omega}(x) 
   = Q^{\cal S}(x, y, s_1, s_2) 
\end{align*}
strongly in $L^2(\R^{2n})$. 

To conclude the proof it remains to show that $Q^{\cal S}_{N}(\cdot, \cdot, s_1, s_2)$ is bounded on compact subsets of $\Omega \times \R^n$. From the corresponding result for the spinless case in Theorem \ref{theo:one-body-matrix} and \eqref{eq:restterm} we see that it is sufficient to prove that $N^{-1} R_N$ is bounded. Since $\operatorname{span}\{\psi_{a_N+1}, \ldots, \psi_N \}$ is a subspace of $\operatorname{span}\{ u_{\alpha} \otimes \delta_{\beta} : a_N + 1 \le \alpha \le \tilde{a}_N, 1 \le \beta \le m \}$, there exists a partial isometry $U = (U_{i,(\alpha,\beta)})$ of this space onto $\operatorname{span}\{\psi_{a_N+1}, \ldots, \psi_N \}$ such that $\psi_i = \sum_{(\alpha,\beta)} U_{i,(\alpha,\beta)} u_{\alpha} \otimes \delta_{\beta}$. We therefore also have the following pointwise estimate: 
\begin{align*} 
  &|R_N(x',x'', s_1, s_2)| \\ 
  &= \bigg| \sum_{i=a_Nm+1}^N \sum_{\alpha = a_N + 1}^{\tilde{a}_N} \sum_{\beta = 1}^m 
        U_{i,(\alpha,\beta)} u_{\alpha}(x') \delta_{\beta}(s_1) 
        \sum_{\alpha' = a_N + 1}^{\tilde{a}_N} \sum_{\beta' = 1}^m 
        \overline{U}_{i,(\alpha',\beta')} \overline{u_{\alpha'}}(x'') \delta_{\beta'}(s_2) \bigg| \\ 
  &\le \left( \sum_{\alpha,\beta} 
    \left| u_{\alpha}(x') \delta_{\beta}(s_1) \right|^2 \right)^{1/2}  
    \left( \sum_{\alpha,\beta} 
    \left| u_{\alpha}(x'') \delta_{\beta}(s_2) \right|^2 \right)^{1/2} \\ 
  &= \left( \sum_{\alpha = a_N + 1}^{\tilde{a}_N} \left| u_{\alpha}(x') \right|^2 \right)^{1/2}  
    \left( \sum_{\alpha' = a_N + 1}^{\tilde{a}_N} \left| u_{\alpha}(x'') \right|^2 \right)^{1/2}.  
\end{align*} 
With the help of Theorem \ref{theo:one-body-conv} we now see that 
\begin{align*} 
  |N^{-1} R(x+ y_N, x-y_N, s_1, s_2)| 
  \le \frac{\tilde{a}_N}{N} \tilde{a}_N^{-1} 
  \sum_{\alpha = a_N + 1}^{\tilde{a}_N} | u_{\alpha}(x+y_N) |^2 + | u_{\alpha}(x-y_N) |^2
\end{align*}
remains bounded  on compact subsets of $\Omega \times \R^n$. \eop 

The proofs of Theorems \ref{theo:wigner-seitz-spin}, \ref{theo:one-body-conv-spin} and \ref{theo:Ex-LDA} are now immediate: \smallskip

\par\noindent{\em Proof of Theorem \ref{theo:wigner-seitz-spin}. } 
By Theorem \ref{theo:one-body-matrix-spin}, 
\begin{align*} 
  P^{\cal S}_{N}(x, y) 
  \to - \frac{1}{2} \sum_{s \in {\cal S}^2} 
    \left| Q^{\cal S}(x, y, s_1, s_2) \right|^2
  = - \frac{2^{n-1} \Gamma^2(\frac{n}{2} + 1) J_{n/2}^2(p_F |y|)}
     {m |\Omega|^2 (p_F |y|)^{n}} \chi_{\Omega}(x) 
\end{align*}
strongly in $L^1(\R^{2n})$ and boundedly in measure on compact subsets of $\Omega \times \R^n$. \eop 

\par\noindent{\em Proof of Theorem \ref{theo:one-body-conv-spin}. } 
The proof follows along the lines of the proof of Theorem \ref{theo:one-body-conv}. It suffices to show that $\rho^{\cal S}_{1, N} \to \bar{\rho}$ in $L^1(\R^n)$. To see this, note firstly that $Q^{\cal S}_{N}(x + y_N, y, s_1, s_2)$ with $y_N = \frac{N^{-1/n} y}{2}$ converges to $Q^{\cal S}(x, y, s_1, s_2)$ strongly in $L^2$ by Theorem \ref{theo:one-body-matrix-spin}, secondly that 
\begin{align*}
  \sum_{s_1,s_2 \in {\cal S}} \int_{\R^n} \left| Q^{\cal S}_{N}(x + y_N, y, s_1, s_2) \right|^2 \, dy
  = \rho^{\cal S}_{1,N}(x)
\end{align*} 
and thirdly that $Q^{\cal S}(x, y, s_1, s_2) = m^{-1} Q(x, m^{-1/n} y) \delta_{s_1s_2}$, whence 
$$ \sum_{s_1,s_2 \in {\cal S}} \int_{\R^n} \left| Q^{\cal S}(x, y, s_1, s_2) \right|^2 \, dy 
   = m^{-1} \int_{\R^n} \left| Q(x, m^{-1/n} y, s_1, s_2) \right|^2 \, dy 
   = \bar{\rho}. $$
\eop

\par\noindent{\em Proof of Theorem \ref{theo:Ex-LDA}.} Let $A \subset\subset \Omega$. By Theorem \ref{theo:one-body-conv-spin}, $\rho^{\cal S}_{1,N}$ converges to $\bar{\rho} \equiv |\Omega|^{-1}$ strongly in $L^p_{\rm loc}(\Omega)$ for every $p \in [1, \infty)$ and so 
$$ \lim_{N \to \infty} c_{\rm x} \int_{A} \left( \rho^{\cal S}_{1,N} \right)^{4/3} (x) \, dx 
   =   3 \left( \frac{3}{32 \pi m} \right)^{1/3} |\Omega|^{-4/3} |A|. $$ 

On the other hand, $P^{\cal S}_{N}$ converges to $P^{\cal S}$ boundedly in measure on $A \times B_1$ and strongly in $L^1$ on $A \times (\R^n \setminus B_1)$ by Theorem \ref{theo:wigner-seitz-spin}, which shows that  
\begin{align*} 
  \lim_{N \to \infty} \int_{A \times \R^n} \frac{P^{\cal S}_{N}(x, y)}{|y|} \, dx \, dy 
  &= - \int_{\R^n} \frac{2^{n-1} \gamma m^{-1/n} \Gamma^2(\frac{n}{2} + 1) 
       J^2_{n/2}(\gamma m^{-1/n} |y|)}{m |\Omega|^2(\gamma m^{-1/n} |y|)^{n+1}} \, dy \, |A| \\ 
  &= - \frac{n \pi^{1/2} \Gamma^{1/n}(\frac{n}{2} + 1)}{m^{1/n} |\Omega|^{1 + 1/n}} 
     \int_0^{\infty} r^{-2} J^2_{n/2}(r) \, dr \, |A|.
\end{align*} 
With the help of the Schafheitlin formula (see, e.g., \cite[p.\ 403]{Watson}), this integral can be determined explicitly: $\int_0^{\infty} r^{-2} J_{n/2}^2(r) \, dr  = \frac{4}{(n^2 - 1)\pi}$ for $n \ge 2$. We therefore obtain 
\begin{align*} 
  \lim_{N \to \infty} \int_{A \times \R^n} \frac{P^{\cal S}_{N}(x, y)}{|y|} \, dx \, dy 
  &= - \frac{4 n \pi^{1/2} \Gamma^{1/n}(\frac{n}{2} + 1)}{(n^2 - 1)\pi m^{1/n} |\Omega|^{1 + 1/n}} |A| \\ 
  &= - \left(\frac{3}{32 \pi m}\right)^{1/3} |\Omega|^{-4/3} |A| 
\end{align*}
for $n = 3$. 
\eop

\subsection*{Acknowledgments}

I am grateful to Gero Friesecke for having drawn my attention to this problem.

 \typeout{References}

\end{document}